\documentstyle[12pt,epsf]{article}
\def\fun#1#2{\lower3.6pt\vbox{\baselineskip0pt\lineskip.9pt
\ialign{$\mathsurround=0pt#1\hfil##\hfil$\crcr#2\crcr\sim\crcr}}}

\newcommand{\be}{\begin{eqnarray}}
\newcommand{\ee}{\end{eqnarray}}
\newcommand{\bd}{\begin{displaymath}}
\newcommand{\ed}{\end{displaymath}}
\newcommand{\ba}{\begin{array}}
\newcommand{\ea}{\end{array}}
\newcommand{\bt}{\begin{tabular}}
\newcommand{\et}{\end{tabular}}

\input epsf
\newcommand{\grpicture}[1]
{
    \begin{center}
        \epsfxsize=400pt
        \epsfysize=0pt
        \vspace{-5mm}
        \parbox{\epsfxsize}{\epsffile{#1.eps}}
        \vspace{5mm}
    \end{center}
}

\begin{document}
\begin{flushright}
TPI--MINN 98/08 \\
ITEP--TH--25/98
\end{flushright}

\vspace{1cm}

\begin{center}
{\large
QCD at $\theta \sim \pi$.}

\vspace{.5cm}

{\bf A.V. Smilga} 
\footnote{Permanent address: {\it ITEP, B. Cheremushkinskaya 25, Moscow
117218, Russia.}}

\vspace{.5cm}

{\it TPI, 116 Church St. S.E., University of Minnesota, MN
55455, USA.}

\vspace{.5cm}

\end{center}

\vspace{.5cm}

\begin{abstract}
Taking into account the terms $\sim m^2$ in the effective chiral lagrangian
,
we show that, at $\theta = \pi$, the theory with 2 light quarks of equal
mass involves two degenerate vacuum states separated by a barrier. For
$N_f = 3$, the energy barrier between two vacua appears already in the 
leading order in mass. This corresponds to the first order phase transition
at $\theta = \pi$. The surface energy density of the domain wall separating
two different vacua is calculated. In the immediate vicinity of the point 
$\theta = \pi$, two minima of the potential still exist, but one of them
becomes metastable. The probability of the false vacuum decay is estimated.

\end{abstract}

\section{Introduction}
 It is very well known that in $QCD$ with $N_f$ massless quarks, chiral 
symmetry $SU(N_f) \otimes SU(N_f)$ is spontaneously broken down to $SU_V(N_f)$.
(This is an experimental fact for $N_f = 2,3$. Probably, no spontaneous 
chiral symmetry breaking occurs for large enough number of flavors
$N_f \sim 8 - 10$ \cite{LNf} which will be of no concern for us here).
Spontaneous symmetry breaking means that the order parameter $<\bar q^i_R
q^j_L>$ ($i,j$ are flavor indices) can acquire an arbitrary direction in
the flavor space. Massless Goldstone particles appear. If the free quark
masses are not zero, the axial chiral symmetry is broken explicitly and
the minimum of the energy functional corresponds to a particular flavor
orientation of the condensate. Goldstones acquire small masses. In the real
World with $m_u \approx 4\ {\rm MeV}$, $m_d \approx 7\ {\rm MeV}$, $m_s 
\approx 150\ {\rm MeV}$,  and $\theta = 0$, the vacuum state is unique.

It is interesting to study also other variants of the theory with different 
values of masses and $\theta$. It was known for a long time \cite{VV,W}
that in the theory with equal light quark masses and $\theta = \pi$, there are 
two degenerate vacuum states. This is best seen in the framework of the 
effective chiral lagrangian describing only the light pseudogoldstone 
degrees of freedom. In the leading order in mass, the effective potential 
is
 \be
\label{Uchir}
V \ =\ - \Sigma \ {\rm Re} \left[ {\rm Tr} \left\{ {\cal M} e^{i\theta/N_f}
U^\dagger \right\}\right]
  \ee
where $U = \exp\{2i \phi^a t^a /F_\pi\}$ ($\phi^a$ are pseudogoldstone 
fields), ${\cal M}$ is the quark mass matrix and $\Sigma$ is the absolute 
value of the quark condensate.

Suppose $N_f =3$, ${\cal M} = m \hat 1$, and $\theta = 0$. The minimum of 
the energy is achieved at $U = \hat 1$. For $\theta = \pi$, there are two 
different minima with $U = \hat 1$ and $U = e^{2\pi i/3} \hat 1$. They are
separated by the energy barrier. The appearance of two vacuum states 
corresponds to spontaneous breaking of the CP--symmetry by the Dashen 
mechanism \cite{Dashen}. 

The situation is, however, more confusing for $N_f =2$. The trace of a $SU(
2)$ matrix is always real which means that, at ${\cal M} = m \hat 1$ and
$\theta =  \pi$, the potential (\ref{Uchir}) does not depend on $U$ at all.
That would mean that the explicit breaking of the chiral symmetry is absent,
pions are massless at this point and the phase transition would be of the 
second rather than of the first order. We understand, however, that the 
chiral symmetry of the original $QCD$ lagrangian {\it is} broken explicitly
by the quark mass term for all values of $\theta$ including $\theta = \pi$, 
and the situation looks paradoxical.

This question was left undiscussed in the original papers \cite{VV,W}, and
only comparatively recently it was shown how the paradox is resolved
\cite{Creutz}. To this end, one should to take into account the terms of 
order $\sim m^2$ in the effective chiral lagrangian. If doing so, the 
continuous vacuum degeneracy at $\theta = \pi$ is lifted, and we obtain 
again only two vacuum states separated by a barrier.
\footnote{One could, of course, anticipate it. For $\theta = 0$ and in the 
leading order in quark mass, the mass of pions is given by the Gell-Mann 
-- Oakes -- Renner relation $F_\pi^2 M_\pi^2 = (m_u + m_d) \Sigma$. A 
theory with $m_u = m_d$ and $\theta = \pi$ is equivalent to the theory
with $m_u = -m_d$ and  $\theta = 0$ (only $\theta_{\rm phys} = \theta +
{\rm arg ( det} {\cal M})$ is relevant). In that case, the pions seem to stay
massless. However, the Gell-Mann 
-- Oakes -- Renner relation is true only in the leading order in $m_q$. 
Higher order corrections bring about a nonzero mass to pions. }

In what follows, we confirm this finding. The main aim of the paper is to 
bring the analysis of Ref.\cite{Creutz} into contact with standard chiral
theory notations and wisdom and to perform some quantitative estimates both
for $N_f = 3$ and $N_f = 2$ for 
the height of the energy barrier, the surface energy density of the domain walls
interpolating between two degenerate vacua at $\theta  = \pi$, and for the 
decay rate of metastable vacuum states at the vicinity of the phase transition 
point.

\section{Three flavors}

Let us discuss first in some details the case $N_f = 3$ where no 
complications due to higher-order terms arise. By a conjugation
$U \to VUV^\dagger$, any unitary matrix $U$ can be brought into the diagonal 
form $U = {\rm diag}
(e^{i\alpha}, e^{i\beta}, e^{-i(\alpha + \beta)} )$. When ${\cal M} = m
\hat{1}$, a conjugation does not change the potential 
(\ref{Uchir}). For diagonal $U$, the latter acquires the form
 \be
 \label{Uab}
U(\alpha, \beta) \ =\ -m\Sigma \left[ \cos \left( \alpha - \frac \theta 3
\right) + \cos \left( \beta - \frac \theta 3
\right) + \cos \left( \alpha + \beta + \frac \theta 3
\right) \right]
  \ee
 The function $U$ has six  stationary points:
\be
\label{statp}
{\rm \bf I} :\ \alpha = \beta = 0, \ \ 
{\rm \bf II} :\ \alpha = \beta = -\frac {2\pi}3, \ \ 
{\rm \bf III} :\ \alpha = \beta =  \frac {2\pi}3 \nonumber \\
{\rm \bf IV}: \  \alpha = \beta = -\frac {2\theta}3 + \pi,\ \ 
{\rm \bf IV}a: \  \alpha = -\alpha - \beta =  -\frac {2\theta}3 + \pi, \ \ 
{\rm \bf IV}b: \  \beta = -\alpha - \beta = -\frac {2\theta}3 + \pi
  \ee
The points {\bf IV}a and {\bf IV}b are obtained from from {\bf IV} by Weyl
permutations and their physical properties are the same. Actually, we have
here not 3 distinct stationary points, but the whole 4--dimensional manifold
$SU(3)/[ SU(2) \otimes U(1)]$ of the physically equivalent stationary
points related to each other by  conjugation. The values of the 
potential at the stationary points are
  \be
  \label{En}
E_{\rm \bf I} = -3m\Sigma \cos \frac \theta 3,\ \ 
E_{\rm \bf II} = -3m\Sigma \cos  \frac {\theta + 2\pi} 3,\ \ 
E_{\rm \bf III} = -3m\Sigma \cos \frac {\theta - 2\pi} 3,\ \ 
\nonumber \\  E_{\rm \bf IV} = m\Sigma \cos  \theta 
  \ee
Studying the expressions (\ref{En}) and the matrix of the second 
derivatives at $\theta = \pi$, one can readily see that {\it i)} the points 
{\bf I} and
{\bf III} are the degenerate minima; {\it ii)} the point {\bf II} is the 
maximum, and {\it iii)} the points {\bf IV} are  saddle points. When 
$\theta$ is slighly less than $\pi$, {\bf I} is a global minimum
while {\bf III} is still a minimum, but of a local variety. The latter 
coalesces
with all three saddle points at $\theta = \pi/2$. At this point the eigenvalues
of the second derivative matrix pass zero and, at still lower values of 
$\theta$, a metastable minimum does not exist. When we instead make 
$\theta$ larger than $\pi$, the picture is  symmetric, only the 
minima {\bf I} and {\bf III} change their roles. At $\theta = 0$, the picture
is exactly reversed compared to what we had at $\theta = \pi$: 
there is one global minimum {\bf I}, two degenerate maxima {\bf II}
and {\bf III}, and a surface of  saddle points {\bf IV}. 
  
  \begin{figure}
  \grpicture{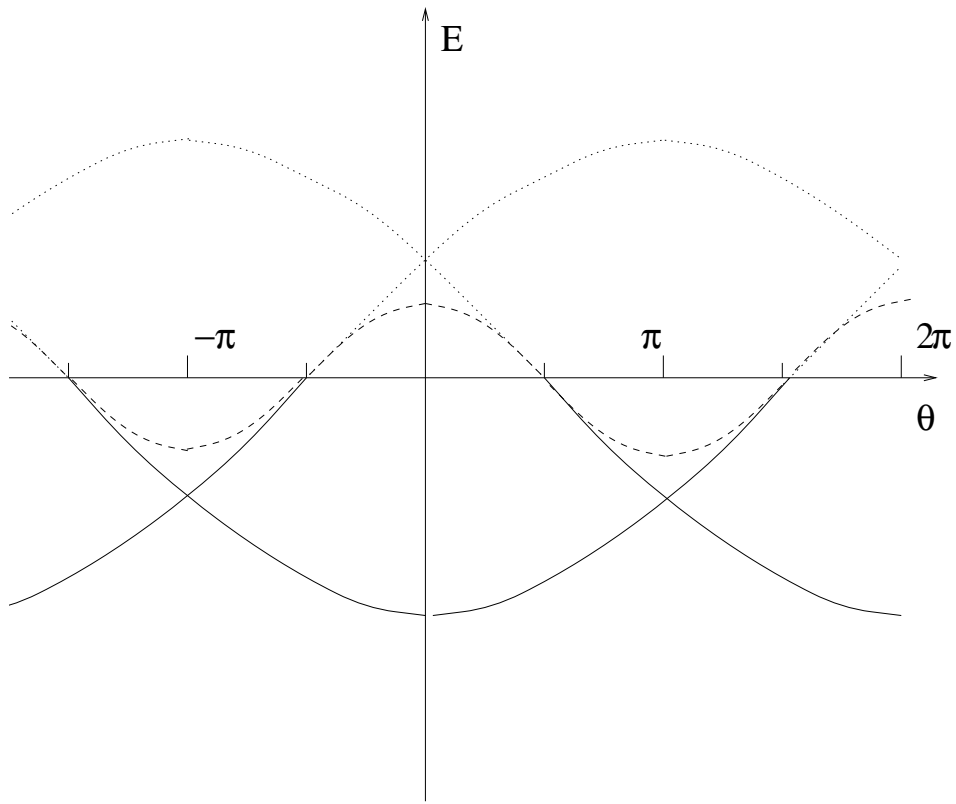}
\caption{Stationary point of $E(\alpha, \beta)$ for different $\theta$. 
Solid lines are the minima, dotted lines are the maxima, and dashed line are 
the saddle points}
  \end{figure}

Fig. 1 illustrates how the stationary points of the potential are moved when
the vacuum angle is changed.
One can show that metastable
vacua are absent at $\theta = 0$ also with physical values of masses.

Two degenerate vacua at $\theta = \pi$ are separated by the domain wall. To
find the profile of this wall, we have to restore the kinetic term 
$\frac {F_\pi^2}4 {\rm Tr} \{\partial_\mu U \partial_\mu U^\dagger \}$ in the 
effective lagrangian and seek for the field configurations depending only
on one spatial coordinate $x$ with the boundary conditions $U(-\infty) = 1
 ,\ \ U(\infty) = e^{2i\pi/3}$ and realizing the minimum of the 
energy functional
  \be
\label{sig}
E\ =\ {\cal A} \sigma \ =\ {\cal A}\int_{-\infty}^\infty  dx \left(
\frac {F_\pi^2}4 {\rm Tr} \{\partial_x U \partial_x U^\dagger \} -
\  m\Sigma {\rm Re} \left[ e^{i\theta/3} {\rm Tr} \{U^\dagger\} \right]
\right)
  \ee
(${\cal A}$ is the total area factor). In our case, it suffices to seek for the 
solutions in the class 
$U = {\rm diag} ( e^{i\alpha(x)}, e^{i\alpha(x)}, e^{-2i\alpha(x)} )$.
Introducing $\gamma = \alpha - \pi/3$, subtracting the vacuum energy and
using the Gell-Mann -- Oakes -- Renner relation, the expression (\ref{sig})
is rewritten as 
  \be
 \label{sigam}
\sigma \ =\ 3F_\pi^2 \int_{-\infty}^\infty dx \left[ \frac 12 \gamma'^2
+ \frac {M_\pi^2}3 \left( \cos \gamma - \frac 12 \right)^2 \right]
  \ee
The corresponding equations of motion with the boundary conditions 
$\gamma(\pm \infty) = \pm \pi/3$ can be readily integrated. The first
integral is
$$\gamma'\  =\ M_\pi \sqrt{2/3} (\cos \gamma - \frac 12)$$
Integrating it further, we obtain the solution
  \be
 \label{solgam}
\cos \gamma \ =\ \frac {E^2 + 4E + 1}{2(E^2 + E + 1)}
  \ee
where $E = \exp \{M_\pi x /\sqrt{2}\}$. The solution (\ref{solgam}) is 
centered at $x = 0$ where it passes through the saddle point {\bf IV}. 
There are, of course, other solutions obtained from Eq.(\ref{solgam}) by a 
shift of $x$. Also we could have chosen the Ans\"atze
$U = {\rm diag} ( e^{i\alpha(x)}, e^{-2i\alpha(x)}, e^{i\alpha(x)} )$
or $U = {\rm diag} ( e^{-2i\alpha(x)}, e^{i\alpha(x)}, e^{i\alpha(x)} )$
and obtain two other wall solutions (with the same properties) passing through 
the saddle points {\bf IV}a and {\bf IV}b.
The wall surface tension is
  \be
\label{sigres}
\sigma \ =\ \frac 9{\sqrt{2}} F_\pi^2 M_\pi \int_0^\infty \frac
{E dE}{(E^2 + E + 1)^2} \ =\ 3\sqrt{2} \left( 1 - \frac \pi {3\sqrt{3}}
\right) M_\pi F_\pi^2
  \ee

Suppose now that $\theta = \pi + \phi$ with $0 < |\phi| \ll 1$. The 
energies of the vacua are not degenerate anymore but are splitted apart by 
the value
 \be
 \label{split}
\Delta {\cal E} \ \approx \ m\Sigma \sqrt{3} |\phi|
 \ee
A metastable vacuum should decay with the formation of bubbles of the
stable phase. The quasiclassical formula for the decay rate per unit time 
per unit volume was derived in \cite{Vol}:
  \be
 \label{rate}
\Gamma \propto \exp \left\{ - \frac {27}2 \pi^2 \frac {\sigma^4}{(\Delta
{\cal E})^3} \right\}
  \ee
where $\sigma$ is the surface tension
the bubble. Substituting here Eqs.(\ref{sigres}, \ref{split}) [ Strictly 
speaking, at $\theta \neq \pi$, the bubble surface tension  does 
not coincide with Eq.(\ref{sigres}) but is somewhat less going to zero at
$\theta = \pi/2$ or $\theta =  3\pi/2$. But at small $|\phi|$, the expression 
(\ref{sigres}) is correct], we obtain
\footnote{Note the difference with the rough estimate 
$\ln \Gamma \sim -F_\pi^2/(M_\pi^2 |\phi|^2)$ for the same quantity in the 
Witten's paper \cite{W}. First, one has $|\phi|^3 $ rather than $|\phi|^2$ 
in the denominator and, second, a huge numerical factor pops up. }
 \be
\Gamma \propto \exp \left\{ - \frac {C F_\pi^2}{M_\pi^2 |\phi|^3} \right\}
 \ee
with
$$ C \ =\ 2^4 \cdot 3^5 \sqrt{3} \pi^2 \left[ 1 - \frac {\pi}{3\sqrt{3}} 
\right]^4 $$
Note that the numerical factor $C$ in the exponent is tremendously large, 
i.e. lifetime of metastable states would be tremendously large (much larger 
than the lifetime of the Universe) for almost 
all $\theta$ in the interval ($\frac \pi 2, \frac {3\pi} 2 $) not too close 
to its boundaries (where metastable states disappear). It is a real pity 
that such a beautiful possibility is not realized in Nature
\footnote{Recently, Halperin and Zhitnitsky argued the existence of 
metastable states in the real $QCD$ at $\theta = 0$ \cite{Zhit}. However, 
their arguments 
were based on a particular model form of the effective potential 
incorporating also glueball degrees of freedom and involving certain cusps.
 The status of this potential is not quite clear by now.}.

\section{Two flavors}
As was mentioned before, we have to take into account here the terms of 
higher order in mass in the effective potential. For $N_f = 2$, it is 
convenient to make benefit of the fact that $SU(2) \otimes SU(2) \equiv O(4)
$ and to use the 4--vector notations so that $U = U_\mu \sigma_\mu = U_0 + 
i U_i \sigma_i, \ \ 
U_\mu^2 = 1$. The 2 $\times$2 complex mass matrix involves 8 real 
parameters which is convenient to "organize" in two different isotopic 4--
vectors
\footnote{We use the notations of Ref.\cite{GL}.}
  \be
\label{hi}
\chi_\mu \ =\ \frac \Sigma{F_\pi^2} 
\left( {\rm Re \ Tr} \{{\cal M} e^{i\theta/2} \},\ 
 {\rm Im \ Tr} \{{\cal M} e^{i\theta/2} \sigma_i \} \right) \nonumber \\
\tilde \chi_\mu \ =\ \frac \Sigma{F_\pi^2} 
\left( {\rm Im \ Tr} \{{\cal M} e^{i\theta/2} \},\ 
 - {\rm Re \ Tr} \{{\cal M} e^{i\theta/2} \sigma_i \} \right)
  \ee
In the second order in $\chi, \tilde \chi$, the most general form of the 
potential is \cite{GL}
  \be
 \label{pot2}
V(U_\mu) \ =\ - F_\pi^2 (\chi_\mu U_\mu) - l_3  (\chi_\mu U_\mu)^2
-  l_7 (\tilde \chi_\mu U_\mu)^2
  \ee
where $l_{3,7}$ are some dimensionless coefficients (the coefficients
 $l_{1,2,4,5,6}$ multiply the structures involving the derivatives of the 
field $U$ in the effective lagrangian). The term $\propto  
  (\chi_\mu U_\mu)  (\tilde \chi_\mu U_\mu)$ is not allowed because it 
would lead to  $CP$ breaking even at $\theta = 0$.

The term $\sim  (\chi_\mu U_\mu)^2$ can always be neglected compared to the 
leading one  for small masses and is not interesting. On the contrary, the
term involving $l_7$ has  a different $\theta$ -- dependence and, for
${\cal M} = m \hat 1$ and $\theta \sim \pi$ when the leading term vanishes, 
determines the whole dynamics.
    
\begin{figure}
  \grpicture{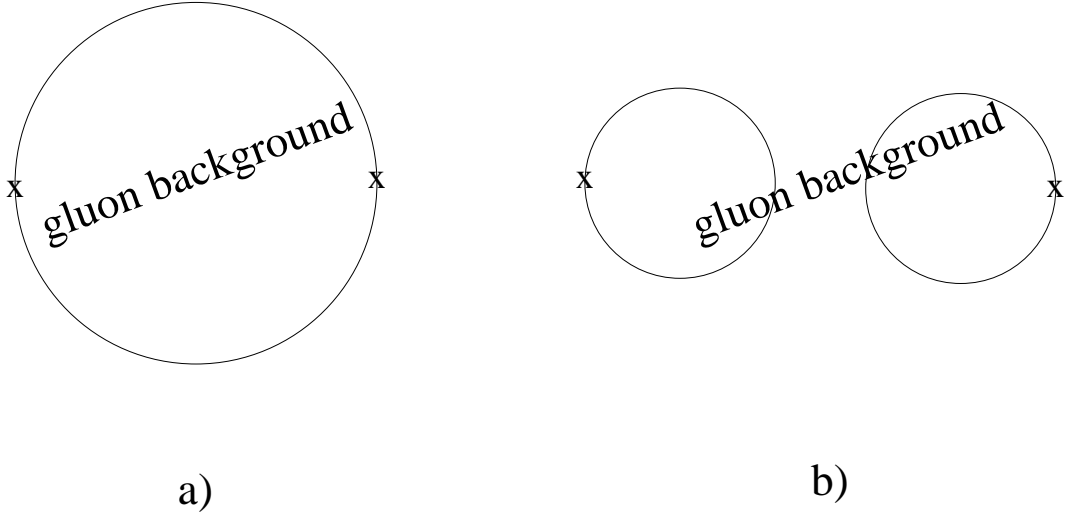}
\caption{Connected and disconnected contributions to the quark current
correlators.}
  \end{figure}
Before proceeding further, let us try to extract an information on the 
numerical value of the constant $l_7$. The following relation belonging to 
the same class as the well--known Weinberg sum rule and derived in
Ref.\cite{GL} is very useful:
  \be
-8\left( \frac \Sigma{F_\pi^2} \right)^2 l_7 \delta^{ik} \ =\ 
\int d^4x \left[ <S^i(x) S^k(0)> - \delta^{ik} <P^0(x) P^0(0)> \right]
 \label{l7SP}
 \ee
where $S^i = \bar q \sigma^i q$ and $P^0 = i \bar q \gamma^5 q$. Let us 
calculate the right--hand side of Eq. (\ref{l7SP}) as an Euclidean 
functional integral. Let us first calculate the correlators in a
{\it particular gauge field background}. Only the connected quark diagram
depicted in Fig. 2a contributes to the correlator of the scalar isovector 
densities. The pseudoscalar correlator receives also a contribution from 
the disconnected graph in Fig. 2b. The solid lines in Fig. 2 stand for the 
quark Green's functions in a particular gauge field background for which we 
use the spectral decomposition
  \be
 \label{specG}
G_A(x,y) \ =\ \sum_n \frac {\psi_n(x) \psi_n^\dagger(y)} {m - i\lambda_n}
 \ee
where $\lambda_n$ are the eigenvalues of the {\it massless} Euclidean Dirac 
operator in an external gauge field and $\psi_n(x)$ are its eigenfunctions. 
All non--zero eigenvalues are paired: for any eigenfunction $\psi_n(x)$ 
with non-zero eigenvalue $\lambda_n$, $\tilde \psi_n(x) = \gamma^5 \psi_n(x
)$ is also an eigenfunction with $\tilde \lambda_n = -\lambda_n$. There are 
also zero modes. For each flavor, their number coincides  with the topological 
charge $\nu = (1/32\pi^2) \int d^4x G_{\mu\nu}^a \tilde G_{\mu\nu}^a $ of the 
gauge field configuration.

Now, for  large Euclidean volumes  $mV\Sigma  \gg 1$, zero mode 
contribution is irrelevant  for the {\it connected} graphs (cf. the 
discussion in Ref.\cite{LS}). On the other hand, the {\it only} 
contribution in the disconnected graph for the pseudoscalar correlator is 
due to zero modes. This contribution is very large $\propto \nu^2/m^2$ and 
is of a paramount importance. Plugging in the Green's functions 
(\ref{specG}) in the correlators in Eq.(\ref{l7SP}), pairing together 
positive and negative $\lambda$, and integrating 
over gauge fields, we obtain for large volumes
  \be
 \label{l7sum}
2\left( \frac \Sigma{F_\pi^2} \right)^2 l_7 \ =\ \frac 1V \left[
\left< \sum'_n \frac{m^2 - \lambda_n^2}{(m^2 + \lambda_n^2)^2} \right>
+ \left< \sum'_n \frac{1}{m^2\ +\ \lambda_n^2} \right>\  -\  
\frac {<\nu^2>}{m^2}  \right]
  \ee
where $\sum'_n$ means the summation over positive  eigenvalues only and the 
symbol $<\ldots>$ stands for the gauge field averaging.
The first term in the R.H.S. comes from the scalar isovector correlator, 
the second term is the contribution of the connected graph to 
the pseudoscalar correlator, and the last term is due to the disconnected 
graph.

Introducing the spectral density
 \be
 \label{dens}
\rho(\lambda) \ =\ \left< \sum_n \delta(\lambda - \lambda_n) \right>\ ,
  \ee
the relation (\ref{l7sum}) is rewritten as
  \be
 \label{l7int}
l_7 \ =\ \frac{F_\pi^4}{\Sigma^2} \left[ m^2 \int_0^\infty \frac 
{\rho(\lambda) 
d\lambda} {(\lambda^2 + m^2)^2}  - \frac {<\nu^2>}{2m^2 V}  \right]
  \ee
This relation belongs to the same class as the famous Banks and Casher 
relation for the fermion condensate
 \be
\label{Banks} 
\Sigma \ =\ 2m\int_0^\infty \frac {\rho(\lambda) d\lambda}
{\lambda^2 + m^2} \ =\ \pi \rho(0) + O(m)
 \ee
For some more examples, see Refs.\cite{Stern,Jac}.

Both terms in (\ref{l7int}) are singular $\propto 1/m$ in the chiral limit. 
The singularity cancels out, however. Indeed, the leading infrared 
contribution in the first term in square brackets  is $
\pi \rho(0)/(4m) \ \sim \Sigma/(4m)$ which is the same as for the second term 
due to the known result for the topological succeptibility in a theory with 
light quarks of the same mass \cite{Creth}
 \be
\frac 1V <\nu^2> \ =\ \frac {m\Sigma}{N_f} + O(m^2)
 \ee
The absense of singularity in the isoscalar pseudoscalar correlator means 
that the corresponding meson is massive: $U(1)$ problem is resolved by the 
't Hooft mechanism due to fermion zero modes in topologically non--trivial 
gauge backgrounds. $l_7$ is given thereby by a constant term $\sim O(1)$ 
which is left out after the cancellation of singular terms. This constant 
is completely determined by the term $\propto m^2$ in $<\nu^2>$. Indeed, 
for $N_f = 2$, the spectral density is analytic at $\lambda = 0$: 
$\rho(\lambda) = \rho(0) + \mu\lambda^2 + \ldots$ \cite{Stern}. The extra 
infrared contribution is of order $O(m)$ and can be neglected.
 \footnote{Note also that the spectral integral in Eq.(\ref{l7int}) 
involves a logarithmic {\it ultraviolet} singularity at large $\lambda$ 
where the spectral density is the same as for free fermions $\rho(\lambda) 
\sim \lambda^3$. It is multiplied, however, by $m^2$ and can be dropped out 
by that reason.}. We finally obtain
  \be
\label{l7nu2}
l_7 \ =\ \lim_{m \to 0}  \frac {F_\pi^4}{4\Sigma^2} 
\left[  \frac{\Sigma \ -\ 2<\nu^2>/(mV)}{m} \right]
  \ee
What can be said about $m$--dependence of the topological succeptibility in 
the next--to--leading order in mass ? Let us first see what happens in a 
theory with two light quarks embedded in the theory  involving also the 
third quark which is much more massive but still
 light enough (like it is the case in the real World). The 
topological succeptibility in the theory with 3 light quarks $m_u = m_d 
\equiv m \ll m_s < \mu_{\rm hadr}$ is given by the expression \cite{VV,W}
 \be
\label{hi3}
\frac {<\nu^2>}{V} \ =\ \frac{m m_s \Sigma}{m + 2m_s} \ =\ \frac {m\Sigma}2
- \frac {m^2 \Sigma}{4m_s} + O(m^3)
 \ee
i.e. the second term is {\it negative}
\footnote{That is quite natural, of course. The presence of an extra light 
quark brings about the suppression of large $\nu$ due to the extra factor 
$m_s^\nu$ in the fermion determinant.} which means that $l_7$ as given by
Eq.(\ref{l7nu2}) is {\it positive}. We obtain
  \be
 \label{l7SU3}
l_7 \ =\ \frac{F_\pi^4}{8m_s \Sigma} \ \equiv \ \frac{F_\pi^2}{6M_\eta^2}
\ \approx \ 5 \cdot 10^{-3}
  \ee
The same result could be obtained in a more direct way if saturating the 
pseudoscalar correlator in Eq.(\ref{l7SP}) by the $\eta$ - meson pole 
\cite{GL}.

The estimate (\ref{l7SU3}) is quite good for the real $QCD$. What we are 
interested in here, however, is a hypothetical theory with $\theta \sim 
\pi$ and just two light flavors. Remarkably, an analytic result for $l_7$
can be obtained also in this case if the number of colors $N_c$ is assumed 
to be large. For large $N_c$, the axial $U(1)$ symmetry is almost not 
affected by the anomaly which means that $\eta'$ -- meson is relatively 
light: $M_{\eta'}^2  \sim \mu_{\rm hadr}^2/N_c$. We can saturate now the 
pseudoscalar correlator by the $\eta'$ pole to obtain
 \be
\label{l7Nc}
l_7 \ =\ \frac {F_\pi^2}{2M_{\eta'}^2}
  \ee
The same result can be obtained via the relation (\ref{l7nu2}). For large 
$N_c$ and $N_f = 2$, the topological succeptibility is known to be 
\cite{RST,W}
  \be
 \label{hiNc}
\frac 1V <\nu^2> \ =\ \frac {m \tau \Sigma }{2\tau + m \Sigma } \ =\ 
\frac {m \Sigma}2 - \frac{m^2 \Sigma^2}{4\tau} + O(m^3)
 \ee
where $\tau$ is the topoligical succeptibility in the pure Yang--Mills 
theory. Again, the term $\sim m^2$ in $<\nu^2>/V$ is negative which leads 
to the positive $l_7$ which coincides with (\ref{l7Nc}) due to the relation
 \be
 \label{Met1}
F_\pi^2 M_{\eta'}^2 \ =\ 4\tau
  \ee
which holds in the limit $m \to 0$.
We assume that $l_7$ is positive also for small number of colors down to 
$N_c = 2$. Indeed, in the limit $m \to \infty$, the topological 
succeptibility should coincide with $\tau$. It is natural that the series 
in $m$ for small masses [the analog of Eq.(\ref{hiNc})] should have alternating 
signs. We cannot, 
unfortunately, formulate this statement (the positiveness of $l_7$)
 as an exact theorem though our {\it suspicion} is that such a theorem can 
somehow be proven.

We are ready now to discuss the vacua dynamics in the region $\theta \sim 
\pi$. Assume $U = {\rm diag} (e^{i\alpha}, e^{-i\alpha})$. The potential
(\ref{pot2}) (with $l_3 = 0$) is 
 \be
 \label{Valth}
V(\alpha) - -2m \Sigma \cos \frac \theta 2 \cos \alpha - 4l_7 m^2 \sin^2 
\frac \theta 2  \left( \frac \Sigma {F_\pi^2} \right)^2 \cos^2 \alpha
 \ee
Defining $\phi = \theta - \pi$, it can be rewritten for small $|\phi|$ as
 \be
\label{Valfi}
V(\alpha) \ =\ m\Sigma \phi \cos \alpha - 4l_7 m^2  \left( \frac \Sigma 
{F_\pi^2} \right)^2 \cos^2 \alpha
 \ee
If
 \be
\label{limfi}
|\phi| < \phi_* = \frac {8l_7 m \Sigma}{F_\pi^4}\ \ ,
 \ee
the function (\ref{Valfi}) has four stationary points:
 \be
 \label{stat2}
{\bf I}\ \ \alpha = 0\ \ \ {\rm with}\ \ \  E_{\bf I} = m\Sigma \phi - 
4l_7 m^2  \left( \frac \Sigma {F_\pi^2} \right)^2 \nonumber \\
 {\bf II} \ \ \alpha = \pi \ \ \ {\rm with} \ \ \ E_{\bf II} = -m\Sigma \phi - 
4l_7 m^2  \left( \frac \Sigma {F_\pi^2} \right)^2 \nonumber \\
{\bf III, \ III}a \ \ \alpha = \pm \arccos \frac {\phi F_\pi^4}{8l_7 m\Sigma} \ \ 
\ {\rm with} \ \ \ E_{\bf III} =  
 \frac  {F_\pi^4 \phi^2}{16l_7} 
 \ee
Again, the points {\bf III}, {\bf III}a are related to each other by
Weyl symmetry and we have actually the whole surface $SU(2)/U(1) \equiv S^2$
of the equivalent stationary points.
Studying the second derivatives $\partial^2V/\partial\alpha^2$ for the 
branches (\ref{stat2}), one readily sees that, in the region (\ref{limfi}),
the points {\bf I} and {\bf II} present local minima (for $\phi < 0$ or
$\theta < \pi$ the absolute minimum is {\bf I} while {\bf II} is a 
metastable state; for $\theta > \pi$, it is the other way round). The points 
{\bf III}  are degenerate maxima. The picture is depicted in Fig. 3. 
Physically, 
it is exactly the same as in the case $N_f = 3$ and we have a first order 
phase transition. Only the width of the region in $\theta$ where two local 
minima coexist is much more narrow than in the case $N_f = 3$ and goes to 
zero in the chiral limit $m \to 0$.

  \begin{figure}
\begin{center}
        \epsfxsize=300pt
        \epsfysize=0pt
        \vspace{-5mm}
        \parbox{\epsfxsize}{\epsffile{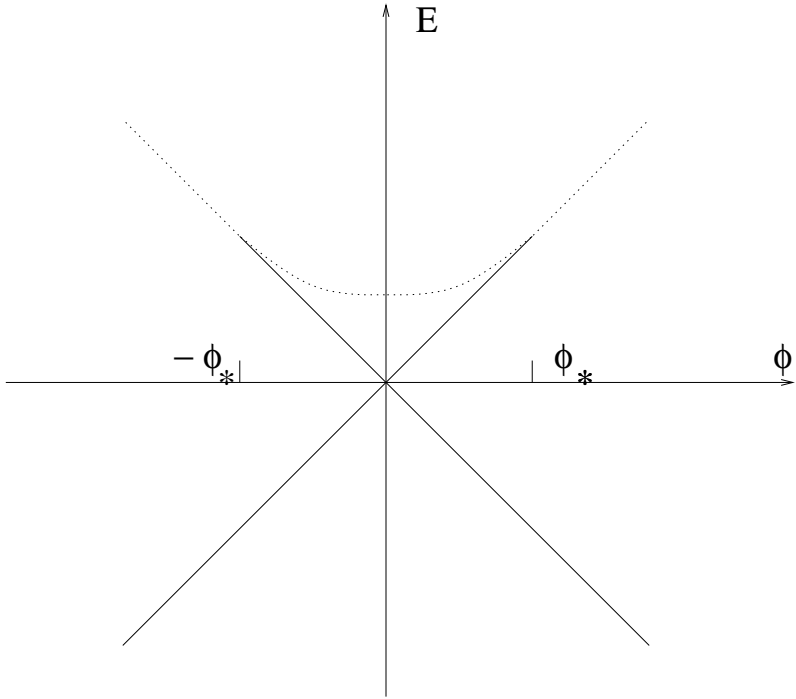}}
        \vspace{5mm}
    \end{center}
\caption{Stationary point of $E(\alpha)$ for $N_f = 2$ 
in the region of small $|\phi| = |\theta - \pi|$. Solid lines are minima 
and dotted lines are maxima.}
  \end{figure}

Note that with negative $l_7$, the picture would be reversed compared to 
that in Fig. 3 so that we would have in the region (\ref{limfi}) 
a surface of degenerate {\it minima} at the 
points {\bf III}. That corresponds to spontaneous breaking of flavor symmetry
$SU_V(2)$ \cite{Creutz}
(when $\theta \neq 0$, the contributions to the partition function
coming from topologically nontrivial sectors are not positively defined,
and the Vafa--Witten theorem \cite{Vafa} which prohibits spontaneous breaking 
of vector flavor
symmetry at $\theta  = 0$ does not apply). Besides, at $ \theta \neq \pi$
we would have the
 maxima of inequal height at the points {\bf I} and
{\bf II}. Instead of one first order phase transition at $\theta = \pi$ we 
would have two consequent second order phase transitions at $\theta = \pi 
\pm \phi_*$. We find this picture  rather unnatural and consider it
 as an additional argument why $l_7$ should be always positive.

It is not difficult now to perform the same program as for $N_f = 3$ and to 
find the surface energy density of the domain wall at $\theta = \pi$ and
the decay rate of metastable vacua at the vicinity of  $\theta = \pi$.
The wall configuration and the energy density at  $\theta = \pi$ are 
obtained by minimizing the functional
 \be
 \sigma \ =\ \int_{-\infty}^\infty \left[\frac {F_\pi^2}2 (\partial_x \alpha)^2
 - 2l_7 m^2 
\left (\frac \Sigma {F_\pi^2} \right)^2 (\cos 2\alpha -1) \right] dx
 \ee
with the boundary conditions $\alpha(-\infty) = 0,\ \ \ \alpha(\infty) = 
\pi$. The  equations of motion have a simple solution
  \be
\alpha(z) \ =\ 2\arctan \left[ \exp \left\{ \sqrt{8l_7} \frac{m\Sigma }
{F_\pi^3} x \right\} \right]
 \ee
The surface energy density is 
 \be
\sigma \ =\ \frac{m\Sigma}{F_\pi} \sqrt{32 l_7}
 \ee
It is much lower numerically than in the case $N_f = 3$ and goes to zero in 
the chiral limit.
The rate of metastable vacuum decay at $|\phi| \ll \phi_*$ is estimated as
 \be
\Gamma \ \propto \ \exp \left\{ -12^3 \pi^2 l_7^2 \frac{m\Sigma}
{F_\pi^4 |\phi|^3} \right\}
 \ee

\section{Schwinger model}

It is very instructive to compare the situation in $QCD_4$ with what happens in
$QED_2$ (the Schwinger model) with two light fermions of equal mass. The 
model was extensively analyzed in Ref.\cite{Coleman}. For zero mass, it 
is exactly solvable. When the mass $m$ is not zero, but much less than 
the gauge coupling constant $g$ (which carries the dimension of mass in two 
dimensions), a systematic expansion in the small parameter $m/g$ (
typically, in some fractional powers thereof) can be built up.

The model is {\it exactly} equivalent to the following bosonized model
 \be
 \label{Lbos}
{\cal L}^{\rm bos} \ =\ \frac 12 (\partial_\mu \phi_+)^2 +
\ \frac 12 (\partial_\mu \phi_-)^2 - \frac {g^2}\pi \left(\phi_+ - \frac
\theta {\sqrt{8\pi}} \right)^2 + C \cos \sqrt{2\pi} \phi_+
 \cos \sqrt{2\pi} \phi_-
 \ee
where $\phi_+$ is the heavy field  and $\phi_-$ 
describes light degrees of freedom. Assuming the
"conformal" normalization for the Euclidean correlator
 \be
\label{norm}
<:\cos \beta \phi_-(x):\ :\cos \beta \phi_-(0):>_{C=0} \ =\ 
\frac 1{2} \left( \frac 1{|x|^{\beta^2/2\pi}} + |x|^{\beta^2/2\pi} \right)
 \ee
and also
 \be
\label{normp}
<:\cos \beta \phi_+(x):>_{C=0} \ =\ 0 ) \ ,
 \ee 
the constant $C$ was calculated to be
\footnote{We have changed the convention for $C$ compared to \cite{crit}
by a factor of 2.}
\cite{crit}
 \be
\label{C}
C \ =\ \frac{mg^{1/2} e^{\gamma/2} 2^{3/4}}{\pi^{5/4}}
  \ee
where $\gamma$ is the Euler constant.
 When $m = 0$, also $C = 0$ so that light and heavy degrees of 
freedom decouple and we have just free massive Schwinger boson with the 
mass
 \be
\label{mubos}
\mu^2 \ =\ \frac{2g^2}{\pi}
  \ee
and a sterile massless particle. When $m \neq 0$, light "quasi-Goldstone"
degrees of freedom acquire mass and start to interact with the heavy ones. 
One can write the effective lagrangian for the light degrees of freedom 
which has largely the same status as the effective chiral lagrangian in 
$QCD$. To lowest order, we can just freeze $\phi_+ = \theta/\sqrt{8\pi}$,
and the lagrangian reads
 \be
\label{Leff0}
{\cal L}_{\rm eff}^0 \ =\ \frac 12 (\partial_\mu \phi_-)^2 + C 
\cos \frac \theta 2\ 
: \cos \sqrt{2\pi} \phi_- :
  \ee
It is nothing else as the Sine-Gordon model. It is exactly solved which 
allows one to find the vacuum energy, the fermion condensate of the 
original theory, and the mass spectrum. Not surprisingly, the lowest states
form an isotopic triplet. A characteric mass scale of these "pions"
\footnote{The only essential difference with the pions of $QCD$ are that 
they are not true goldstones and decouple in the chiral limit. This is
related to the Merman-Wigner-Coleman theorem \cite{MWC} forbidding the 
existence of massless interacting particles  and, thereby,
a spontaneous breaking of a continuous symmetry in two dimensions.}
 is $\sim \left(C \cos \frac \theta 2 \right)^{2/3} \sim \left(m^2 g \cos^2 
\frac \theta 2 \right)^{1/3}$. See 
Ref.\cite{crit} for the exact calculation. 

When $\theta \sim \pi$, the leading term in the effective potential 
vanishes, however, and we are in a position to take into account higher 
order corrections in the Born--Oppenheimer parameter $m/g$. It is 
convenient to rewrite the lagrangian (\ref{Lbos}) in terms of $\chi = \phi_+ -
\ \theta/\sqrt{8\pi}$:
  \be
 \label{Lchi}
{\cal L} \ =\ \frac 12 (\partial_\mu \chi)^2 - \frac {\mu^2}2  \chi^2
+ \frac 12 (\partial_\mu \phi_-)^2  + \nonumber \\
C \left[ \cos \frac \theta 2  \cos \sqrt{2\pi} \chi
- \sin \frac \theta 2    \sin \sqrt{2\pi} \chi \right]
:\cos \sqrt{2\pi} \phi_-:
 \ee
The correction $\sim C^2 \propto m^2$ in the effective potential is given by 
the expression
 \be
\label{DelV}
\Delta V(x) \ =\ -\frac {C^2}2 : \cos \sqrt{8\pi} \phi_-(x): 
\left[ \cos^2 \frac \theta 2 \int d^2y |y|<\cos \sqrt{2\pi} \chi(y)  
\cos \sqrt{2\pi} \chi(0) >_{C=0} \right. \nonumber \\  + \left.
 \sin^2 \frac \theta 2 \int d^2y |y| <\sin \sqrt{2\pi} \chi(y)  
\sin \sqrt{2\pi} \chi(0) >_{C=0} \right]
 \ee
This formula has the same meaning as Eq.(6.30) in the paper by Coleman
\cite{Coleman}. We only 
used accurately the conformal fusion rules
 \be
 \label{fusion}
:e^{i\beta \phi(x)}: \times :e^{i\beta \phi(0)}: \ \ =\ |x|^{\beta^2/2\pi}
:e^{2i\beta \phi(0)}:\ + \ldots \ \ ,
 \ee
neglected the operators of higher dimension in Eq.(\ref{fusion})
(they bring about the corrections of still
higher order in $m$), omitted an irrelevant additive constant, and 
took into account {\it all} loops of the heavy field $\chi(x)$ drawn in  
Fig. 4 (Coleman only took the first graph which is 
much similar to the graph with the $\eta'$ exchange which saturates the 
pseudoscalar correlator and gives the leading contribution in $l_7$ in 
$QCD_4$ in the large $N_c$ limit. It is good enough for a qualitative 
estimate, and, as we will soon see, the account of other graphs only brings 
about a certain numerical factor which is very close to 1.). 

  \begin{figure}
  \grpicture{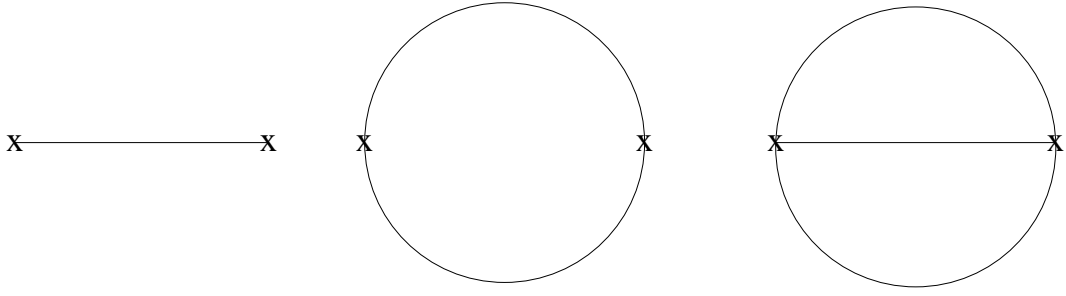}
\caption{Some graphs contributing in the effective lagrangian in the second 
order. Crosses stand for the insertion of light vertex $\cos (\sqrt{2\pi} 
\phi_-)$ and solid lines describe heavy Schwinger bosons. The graphs with 
even number of the lines contribute in $\kappa_0$ and the graphs with odd 
number in $\kappa_1$.}
  \end{figure}

Expanding the 
integrand in $\chi(x)$ and $\chi(0)$, disregarding
the "tadpole" contributions involving the correlators  in coinciding points
[they vanish due to the convention (\ref{normp})], and substituting the free 
massive boson Green's function 
 \be
 \label{Green}
<\chi(x) \chi(0)> \ =\ \int \frac {e^{i {\bf k x}}}{k^2 + \mu^2}
\frac {d^2k}{4\pi^2} \ =\ \frac 1{2\pi} K_0(\mu x)\ ,
 \ee
we finally obtain
  \be
\label{Leff01}
{\cal L}_{\rm eff} \ =\ \frac 12 (\partial_\mu \phi_-)^2 + C 
\cos \frac \theta 2 :\cos \sqrt{2\pi} \phi_-: + \nonumber \\
\frac{\pi^{5/2} C^2}{\sqrt{8}g^3} \left[ \kappa_0 \cos^2 \frac \theta 2
+  \kappa_1 \sin^2 \frac \theta 2 \right] :\cos \sqrt{8\pi} \phi_-:
 + O(C^3)
  \ee
where
 \be
\label{kappa}
\kappa_0 \ =\ \int_0^\infty z^2 dz \{\cosh [K_0(z)] - 1 \}\ = \ .163\ldots
\nonumber \\
\kappa_1 \ =\ \int_0^\infty z^2 dz \sinh [K_0(z)] \ = \ 1.604\ldots
  \ee
The constant $\kappa_0$ is an exact counterpart of the constant $l_3$ in 
the effective potential (\ref{pot2}) for $QCD_4$. The constant $\kappa_1$
is  an exact counterpart of $l_7$, the quantity of primary interest for us 
here. We see that $\kappa_1$ is positive (and that serves as an additional 
argument why $l_7$ should be positive in $QCD_4$). Note that if plugging
in just $K_0(z)$ instead of $\sinh[K_0(z)]$ in the integral (\ref{kappa}) 
for $\kappa_1$ ({\it that} corresponds to taking into account only the 
graph with one heavy particle exchange), we would obtain $\kappa_1 = \pi/2$
instead of 1.604. The effect of higher loops increases the value of 
$\kappa_1$ just by $\sim$ 3\%. 

Even the analog of the 
domain wall at $\theta  = \pi$ exists in this nice model. In two 
dimensions, a domain wall is just a particle. On the classical level, it is a 
Sine--Gordon soliton corresponding to interpolating between the points
$\phi_- = 0$ and $\phi_- = \sqrt{\pi/2}$. Soliton and antisoliton form 
an isotopic doublet.

It would be very interesting  to analyze the quantum problem and to find the 
mass spectrum and other characteristics of the model   to the order $\sim 
C^2$. Our impression is that it is not so easy to 
do  at $\theta = \pi$ :
the problem is that $\beta = \sqrt{8\pi}$ is  exactly a boundary value of 
the coupling. The Sine--Gordon theory with $\beta > \sqrt{8\pi}$ is sick:
the hamiltonian does not have a ground state etc. This displays itself in 
the fact that the factor multiplying $:\cos \sqrt{8\pi}\phi_- :$ in 
Eq.(\ref{Leff01}) is 
dimensionless and we do not have a mass parameter out of which
the soliton mass could be
 composed. Probably,
at $\theta = \pi$, the Born--Oppenheimer expansion breaks down in this case
and one has to analyze the full lagrangian (\ref{Lbos}) by approximate methods
(cf. recent \cite{Hoso}). The lagrangian 
(\ref{Leff01}) may  still be taken at the face value at other values of $\theta$
and could allow to find corrections in mass to the exact results of Ref.
\cite{crit} at $\theta = 0$ etc.
 But this is beyond the scope of the present paper.

\vspace{.2cm}

{\bf Acknowledgements}: I am indebted to M. Creutz, H. Leutwyler and
M. Shifman for illuminating discussions.  This work was supported in part by 
the RFBR--INTAS grants 
93-0283 and 94-2851, by the RFFI grant 97-02-16131, by the U.S. Civilian 
Research and Development Foundation under award \# RP2-132, and by the 
Schweizerischer National Fonds grant \# 7SUPJ048716.

\end{document}